\documentclass[12pt,a4paper]{article}
\usepackage{epsfig}
\usepackage{graphicx}                  
\usepackage{ae}
\usepackage{amsmath}
\usepackage{amssymb}
\usepackage{graphics}
\usepackage{dcolumn}
\usepackage{bm}
\begin{document}
\begin{center}

{\bf \Large Building of a 4-channel TTL scaler for counting detector signals}\\
\vspace{0.1cm}
S.~Sahu$^a$,
R.~P.~Adak$^b$,
S.~Biswas$^{b*}$,
T.~Mishra$^a$,
D.~Nag$^b$,
R.~N.~Patra$^c$,
S.~Rudra$^d$,
P.~K.~Sahu$^a$,
and S.~Swain$^a$\\

$^a$ Institute of Physics, Sachivalaya Marg, P.O: Sainik School, Bhubaneswar - 751 005, Odisha, India\\
$^b$ Bose Institute, Department of Physics and Centre for Astroparticle Physics and Space Science
(CAPSS), EN-80, Sector V, Kolkata-700091, India\\
$^c$ Variable Energy Cyclotron Centre, 1/AF Bidhan Nagar, Kolkata-700 064, West Bengal, India\\
$^d$ Department of Applied Physics, University of Calcutta, 92, APC Road, Kolkata-700 009, West Bengal, India\\
$^*$E-mail: saikat@jcbose.ac.in, saikat.ino@gmail.com, saikat.biswas@cern.ch
\end{center}

\abstract{A scaler has been fabricated to count the signals from any radiation detector. It can count signals of frequency up to 140~kHz. Transistor Transistor Logic (TTL) is used in this scaler. In this article the details of the design, fabrication and operation processes of the scaler is presented. }

\section{Introduction}\label{intro}
In India R\&D projects on various gaseous detectors such as Gas Electron Multiplier (GEM), Resistive Plate Chamber (RPC), Single Wire Proportional Chamber (SWPC), Multi Wire Proportional Chamber (MWPC) and scintillator detectors are being carried out now a days for various low energy nuclear physics and high energy physics (HEP) experiments \cite{RNP, KKM, RG, SR14, APN}. During the characterisation of any detector, the analog input signals from the detector are fed into discriminator and the corresponding digital signals are counted. Different scalers or counters are commercially available for this purpose. The available commercial scalers are quite expensive. An effort is going on in our collaborative work to build a cheap in house scaler. In that spirit  a rising edge triggered 4-channel  TTL (Transistor Transistor Logic) scaler has been developed to record the number of pulses in a given interval of time. The four channels are independent and each channel is capable of capturing maximum 4,294,967,295~(2$^{32}$-1) number of pulses i.e. each channel can count maximum 4,294,967,295~(2$^{32}$-1) number of signals. In this report the details of the design, fabrication and calibration of the scaler is presented.

\section{Design principle}\label{construct}

\begin{figure}[htb!]
\begin{center}
\includegraphics[scale=0.35]{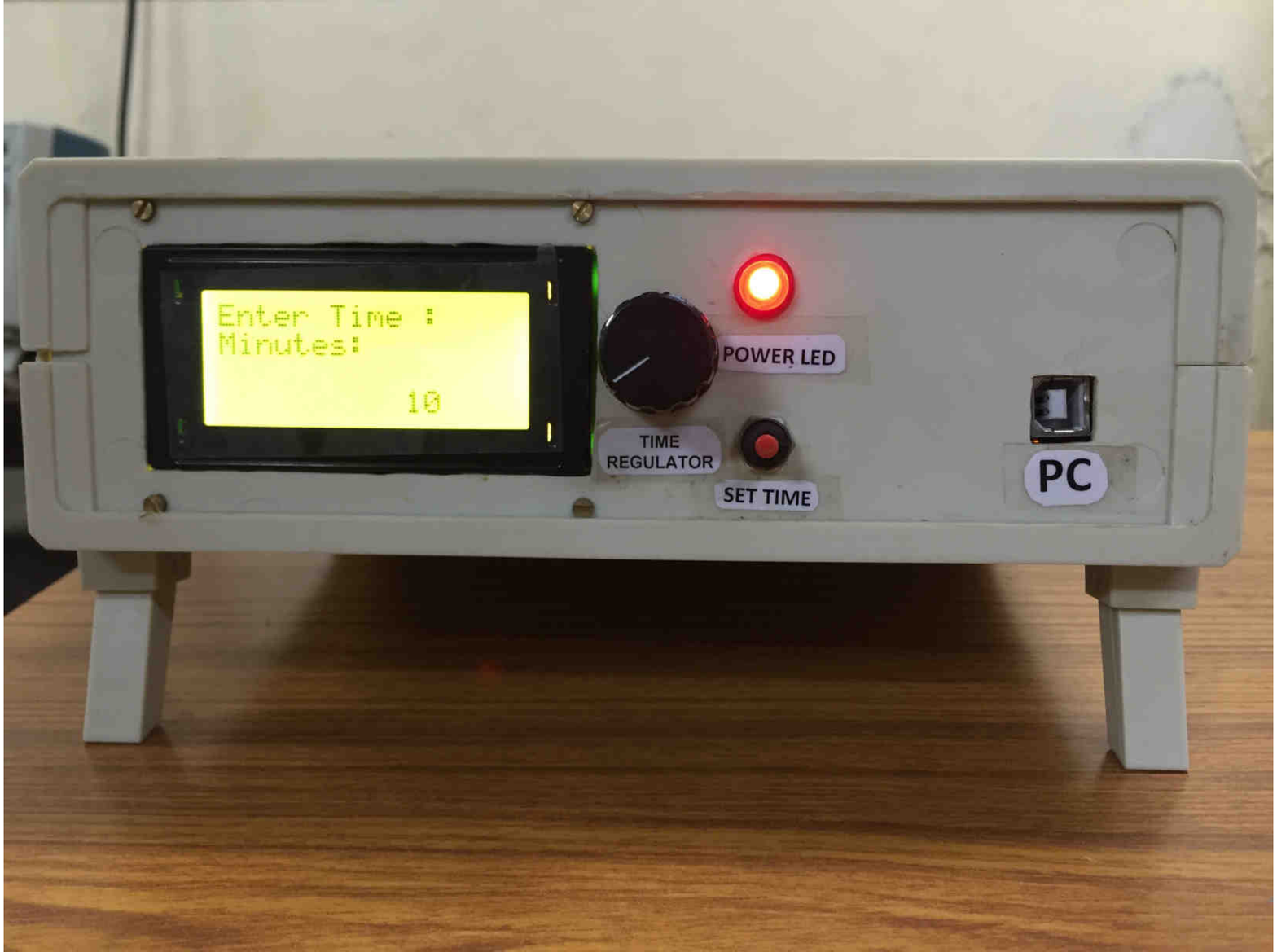}
\caption{\label{Front} Front side view of the scaler.}\label{Front}
\end{center}
\end{figure}
\begin{figure}[htb!]
\begin{center}
\includegraphics[scale=0.35]{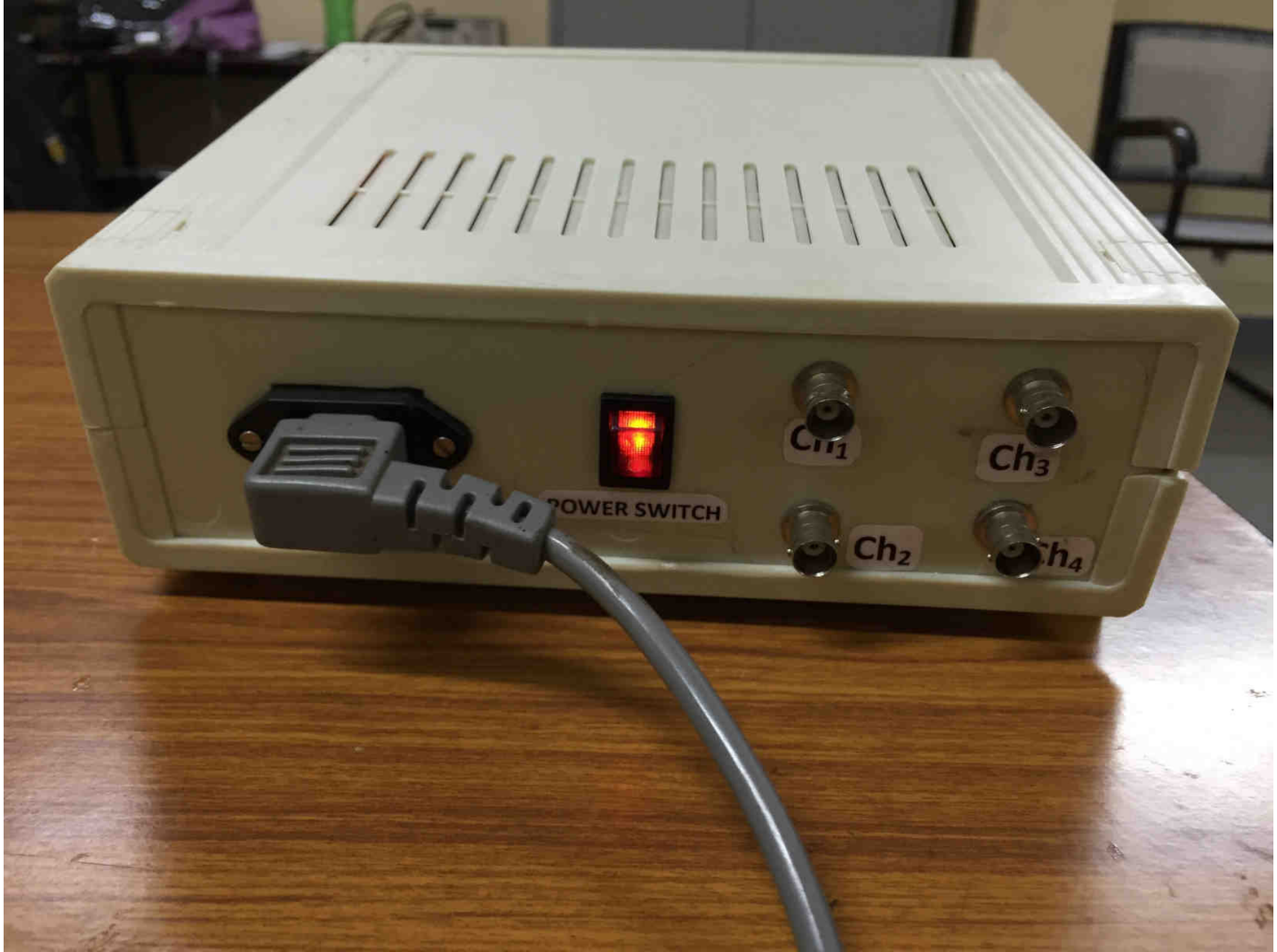}
\caption{\label{Back} Back side view of the scaler.}\label{Back}
\end{center}
\end{figure}
One standalone 4-channel edge triggered TTL scaler is designed here to count the pulses from the detector \cite{SS}. The scaler is powered directly from the 220~V AC line. The front and back side view of the scaler are shown in Fig.~\ref{Front} and Fig.~\ref{Back} respectively. The power switch of the scaler is placed at the rare side of the device as shown in Fig.~\ref{Back}. There is a red LED which will glow when the scaler is switched on. There are four inputs to the scaler. All the inputs are fed with BNC connector as shown in Fig.~\ref{Back}. The analogue signals from the detector is first discriminated from a discriminator module and the TTL signal is generated either from the SCA or from a NIM-TTL converter. The TTL signal is fed to one of the inputs of the scaler. The scaler can accept user command by external knob (potentiometer) for setting up of sampling time. The maximum sampling time can be set to 120 minutes. The black knob is marked as "Time Regulator" in  Fig.~\ref{Front}. After setting up of sampling time one pulse switch (KEY) is pressed to start the count. This switch is marked as the "Set Time" in Fig.~\ref{Front}. The sampling time is set in minutes as shown in Fig.~\ref{Display}. The elapsed time during counting are displayed on the LCD in seconds as shown in Fig.~\ref{Display2}. At the end of counting the number is displayed on the LCD as shown in Fig.~\ref{Display3}. Each channel has 10 digit display as seen in Fig.~\ref{Display3}. The scaler can be controlled by computer and the counted numbers can automatically be stored in the computer.

\begin{figure}[htb!]
\begin{center}
\includegraphics[scale=0.25]{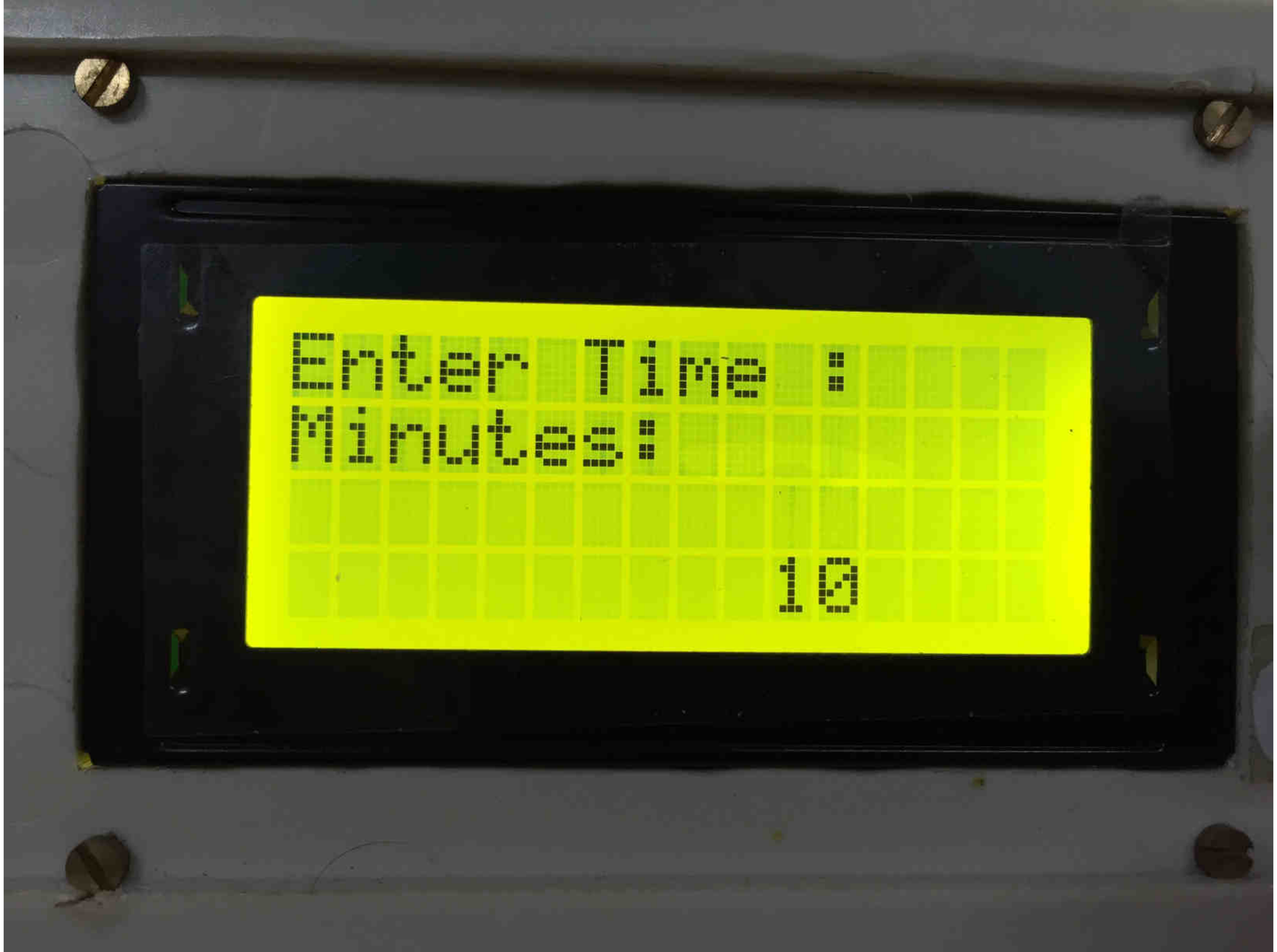}
\caption{\label{Display} Display unit.}\label{Display}
\end{center}
\end{figure}
\begin{figure}[htb!]
\begin{center}
\includegraphics[scale=0.25]{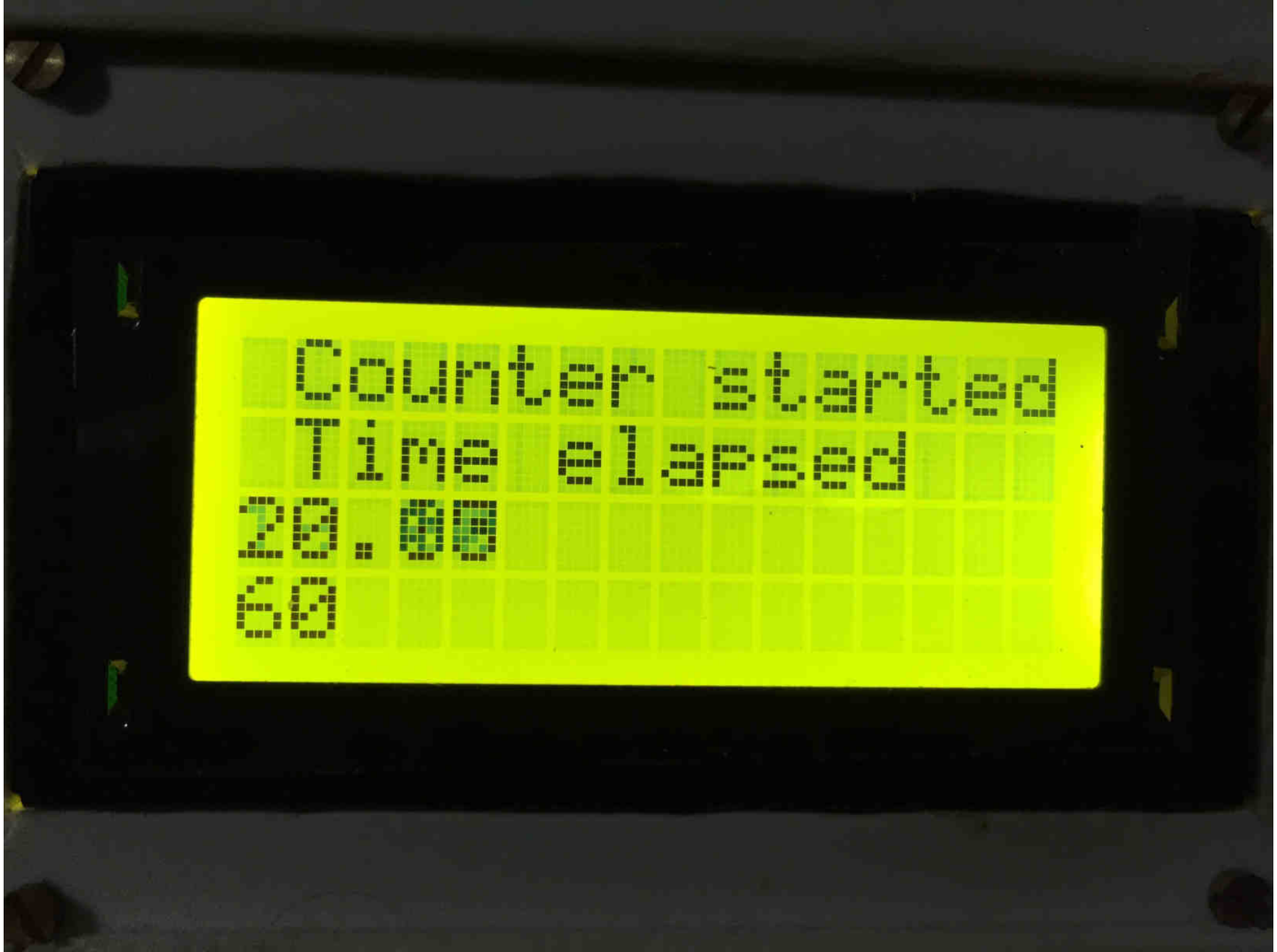}
\caption{\label{Display2} Display during run time.}\label{Display2}
\end{center}
\end{figure}
\begin{figure}[htb!]
\begin{center}
\includegraphics[scale=0.25]{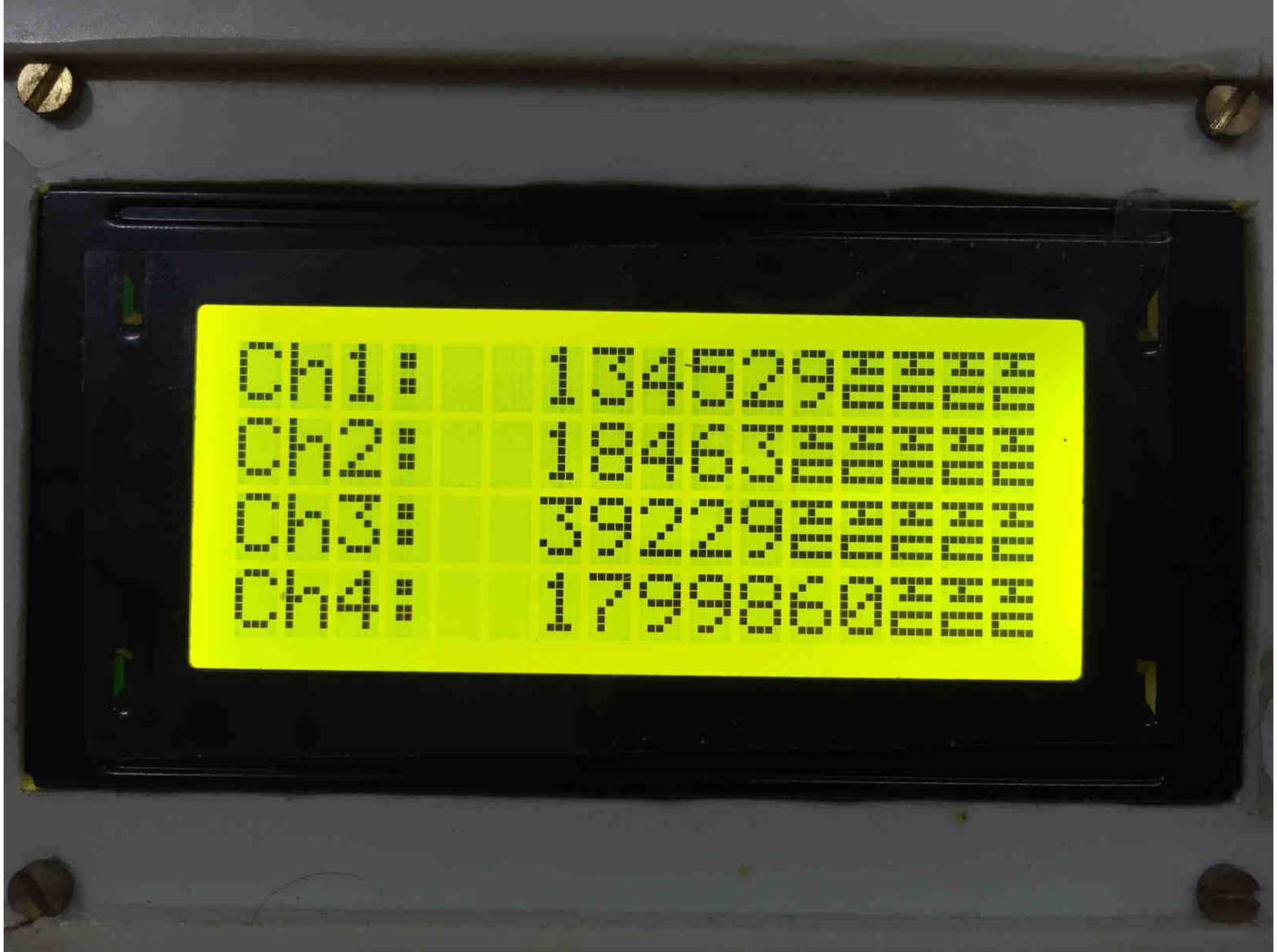}
\caption{\label{Display3} Display at the end of counting.}\label{Display3}
\end{center}
\end{figure}
\begin{figure}[htb!]
\begin{center}
\includegraphics[scale=0.35]{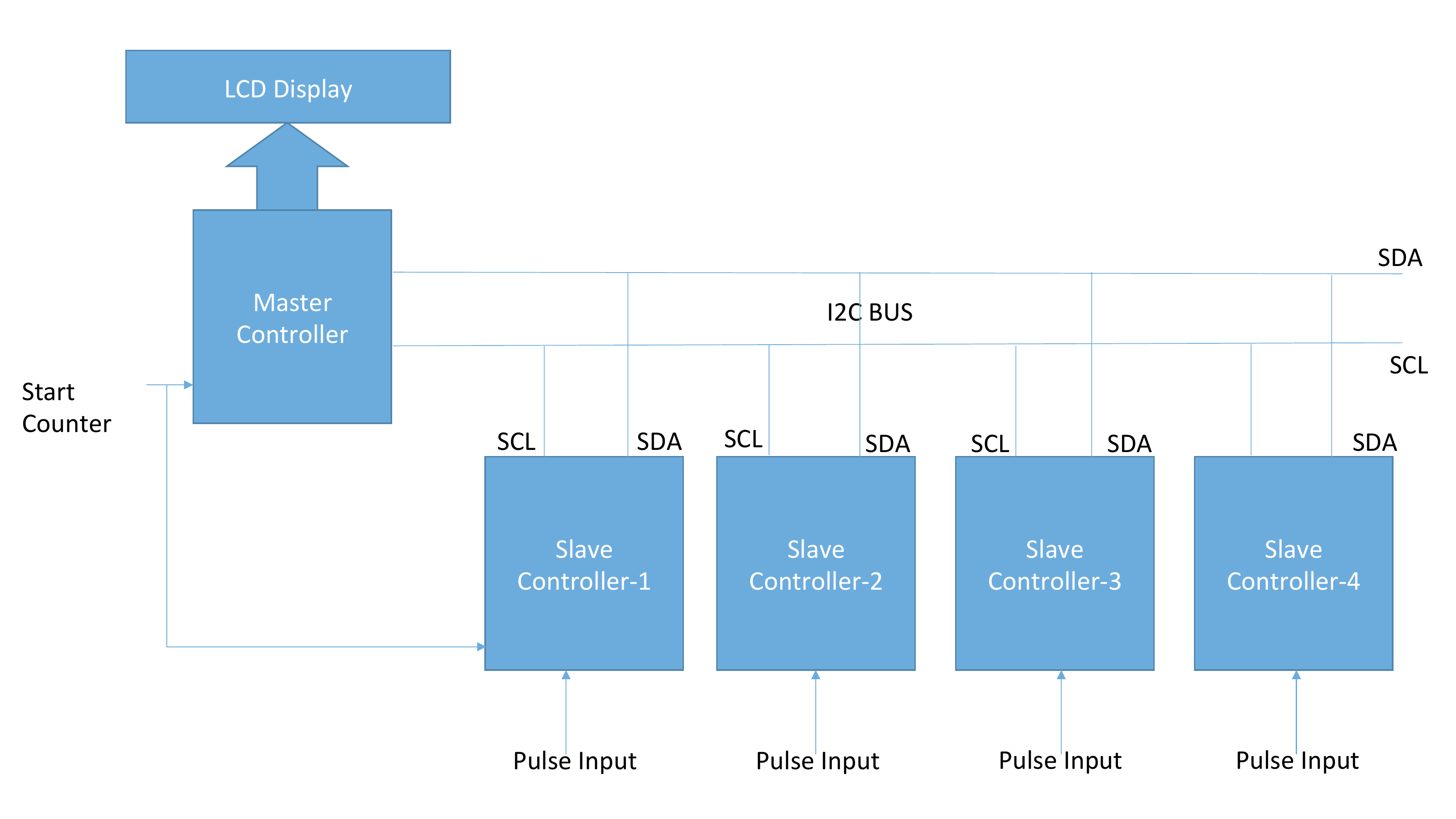}
\caption{\label{block} The block diagram of the scaler.}\label{block}
\end{center}
\end{figure}
\begin{figure}[htb!]
\begin{center}
\includegraphics[scale=0.6]{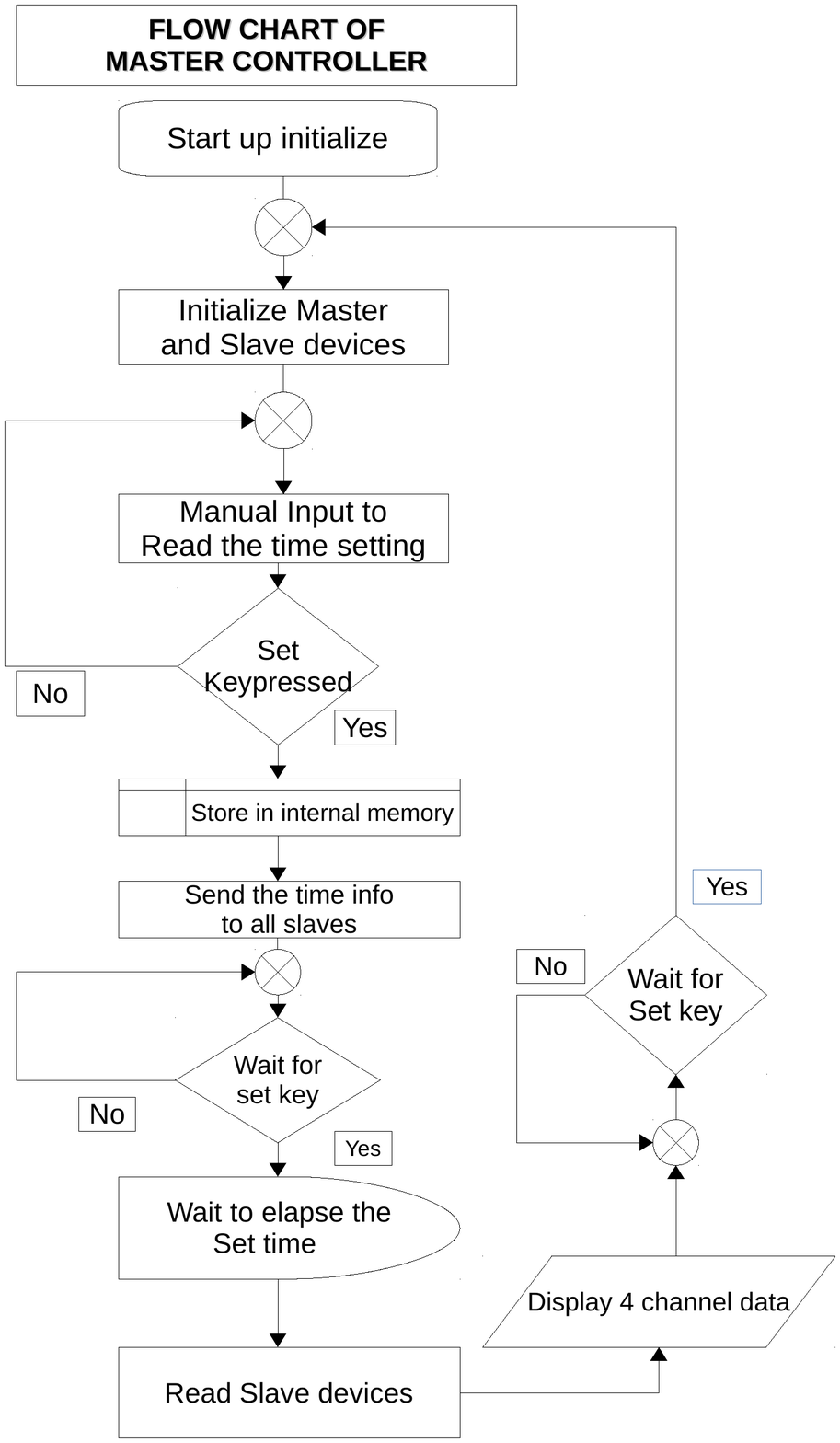}
\caption{\label{master} The flowchart of master controller.}\label{master}
\end{center}
\end{figure}
\begin{figure}[htb!]
\begin{center}
\includegraphics[scale=0.6]{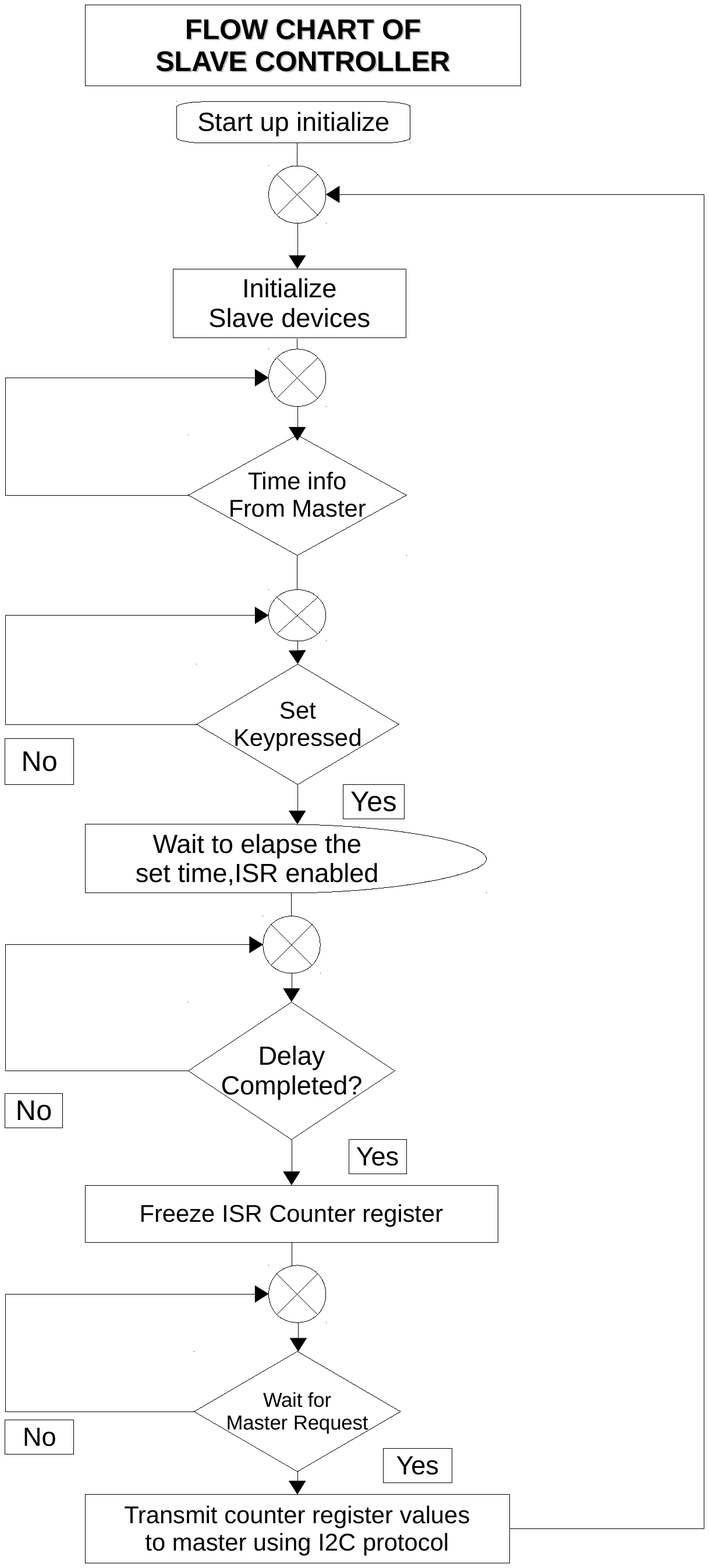}
\caption{\label{slave} The flowchart of slave controller.}\label{slave}
\end{center}
\end{figure}
\begin{figure}[htb!]
\begin{center}
\includegraphics[scale=0.5]{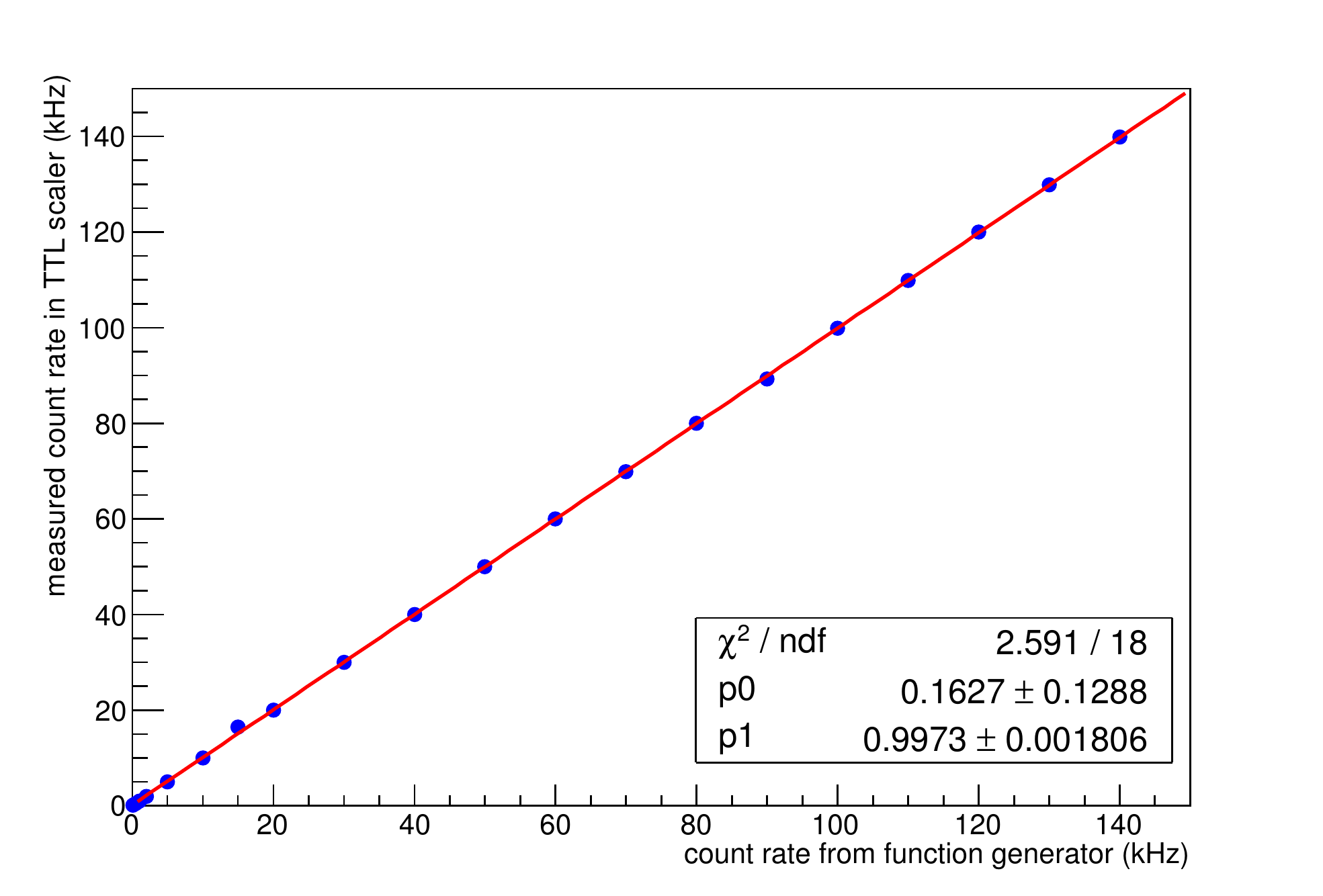}
\caption{\label{calib} The calibration curve.}\label{calib}
\end{center}
\end{figure}

The scaler has been designed to avoid multiple counting with larger pulse width. So, the edge trigger with fast response has been designed i.e. the scaler counts only when the digital signal changes its state from 0 to 1. A rate of maximum 140 kHz can be counted without any delay independently in four channels. The time is calculated in millisecond accuracy so a high precession counting can be achieved. The entire scaler is designed using independent Atmega328 microcontrollers. There are  five numbers of  microcontrollers out of them one is MASTER and other four are SLAVES and the MASTER-SLAVE communicates with (Inter-Integrated Circuit)~I$^2$C protocol. The block diagram of the scaler is shown in Fig.~\ref{block}. The functions of the Master and Slave controllers are given as flow chart in Fig.~\ref{master} and Fig.~\ref{slave} respectively.

\section{Calibration of the TTL scaler}\label{calibration}
The TTL scaler is calibrated using a function generator. The function generator can provide different kind of signals of different frequencies. In this work square wave signal is considered. The amplitude of the square wave signal is kept constant and the frequency is varied from a few Hz to a few kHz. The frequency of the output signals from the function generator can be obtained from the function generator itself. That square wave signal of known frequency is fed to the input of the scaler. The number of signals are counted for 2 minutes for each settings and from it the rate is calculated. This calibration is done for all four channels individually and the calibration curve is drawn. The calibration curve for the count rate of one channel is shown in Fig.~\ref{calib}. The curve is found to be a straight line up to 140 kHz, with a slope of 0.99 and intercept $\sim$~0.16. Same parameters are obtained for other three channels also. 

\section{Summary and outlooks}
One 4-channel TTL scaler has been fabricated to count the signals from any detector. This is a part of our regular detector R\&D programme. The scaler has the following features. (a) The scaler has 4 channels, (b) each channel has 10 digit display, (c) the scaler can accept TTL input, (d) it can accept the maximum count rate of 140 kHz, (e) the maximum preset time can be set to 120 minutes and (f) count is displayed once the counting is stopped. The count rate of the TTL scaler is calibrated with a function generator. The calibration curve is found to be a straight line with a calibration factor of 0.99 and intercept $\sim$~0.16. So this scalar is a low priced one and suitable for the counting of signals from a detector.

In future we are planning to build a scaler in standard NIM format and will be communicated at a later stage.

\section{Acknowledgements}
We would like to acknowledge Prof. S.~Raha, Prof. S.~K.~Ghosh, Dr. S.~Das, Dr. R.~Ray of Bose Institute, Kolkata, Prof. S.~Panda of IOP, Bhubanswar and Dr. S.~Chattopadhyay, Dr. T.~K.~Nayak of VECC, Kolkata for  their support during course of this work. We would also like to thank the RD51 collaboration for helping us and giving valuable suggestions in the course of our different works.


\noindent
\begin{thebibliography}{50}

\bibitem{RNP} R.N. Patra et al., Nuclear Instruments and Methods in Physics Research A 824 (2016) 501, [arXiv:1505.07768].
\bibitem{KKM} K.K. Meghna, et al., Nuclear Instruments and Methods in Physics Research A 816 (2016) 1-8.
\bibitem{RG} Rajesh Ganai et al., 2016 JINST 11 P04026, doi: 10.1088/1748-0221/11/04/P04026 [arXiv:1510.02028].
\bibitem{SR14} Sharmili Rudra et al., Proc. of the DAE Symp. on Nucl. Phys, Vol. 59, (2014), 870-871. 
\bibitem{APN} A.P. Nandan, et al., Nuclear Instruments and Methods in Physics Research A 824 (2016) 606, [arXiv:1407.7181].






\bibitem{SS} S. Sahu et al., Proceedings of the DAE Symp. on Nucl. Phys. 60 (2015) 958.


\end{thebibliography}
\end{document}